\documentclass[aps,prb,twocolumn,a4,showpacs,superscriptaddress]{revtex4-1} 
\usepackage{amsmath,amssymb,amsfonts,upgreek}
\usepackage{mathtools}

\usepackage[english]{babel}
\usepackage[utf8x]{inputenc}
\usepackage[T1]{fontenc}
\usepackage{bm}
\usepackage{nicefrac}
\usepackage{siunitx}
\usepackage{textcomp}

\usepackage{rotating}
\usepackage[colorinlistoftodos]{todonotes}
\usepackage[colorlinks=true, allcolors=blue]{hyperref}
\usepackage{relsize}

\usepackage{amsmath,amssymb,amsfonts,upgreek}
\usepackage{gensymb}
\usepackage{color}
\usepackage{graphicx}
\usepackage{setspace}
\usepackage{multirow}
\usepackage{hhline}
\usepackage{bm}
\usepackage{comment}
\usepackage{natbib}

\begin{document}
\title{Anomalous thermoelectric transport phenomena from interband electron-phonon scattering}

\author{Natalya S.\ Fedorova}
\email{natalya.fedorova@list.lu}
\affiliation{John A. Paulson School of Engineering and Applied Sciences, Harvard University, 29 Oxford Street, Cambridge, MA 02138, USA}
\affiliation{Ferroic Materials for Transducers group, Materials Research and Technology Department, Luxembourg Institute of Science and Technology,
5 Avenue des Hauts-Fourneaux, L-4362 Esch/Alzette, Luxembourg}

\author{Andrea Cepellotti}
\affiliation{John A. Paulson School of Engineering and Applied Sciences, Harvard University, 29 Oxford Street, Cambridge, MA 02138, USA}

\author{Boris Kozinsky}
\email{bkoz@seas.harvard.edu}
\affiliation{John A. Paulson School of Engineering and Applied Sciences, Harvard University, 29 Oxford Street, Cambridge, MA 02138, USA}
\affiliation{Robert Bosch LLC Research and Technology Center, Cambridge, MA 02139, USA}

\begin{abstract}
The Seebeck coefficient and electrical conductivity are two critical quantities to optimize simultaneously in designing thermoelectric materials, and they are determined by the dynamics of carrier scattering. We uncover a new regime where the co-existence at the Fermi level of multiple bands with different effective masses leads to strongly energy-dependent carrier lifetimes due to intrinsic electron-phonon scattering. In this anomalous regime, electrical conductivity decreases with carrier concentration, Seebeck coefficient reverses sign even at high doping, and power factor exhibits an unusual second peak. We discuss the origin and magnitude of this effect using first-principles Boltzmann transport calculations and simplified models. We also identify general design rules for using this paradigm to engineer enhanced performance in thermoelectric materials.
\end{abstract}

\maketitle

\section{Introduction}
\label{sec::intro}
Thermoelectric materials hold promise as a basis of solid-state devices for capturing waste heat from industrial processes and automotive engines  \cite{bell2008cooling, chen201recent3, mao2018advances,snyder2008complex}. The efficiency of thermoelectrics is characterized by the figure of merit $ZT=\sigma S^2 T/\kappa$, where $\sigma$ is the electrical conductivity, $S$ - the Seebeck coefficient, $T$ - the absolute temperature, and $\kappa$ - the thermal conductivity. Strategies for increasing the figure of merit are being actively researched, using experiments and computation, and they typically aim to maximize the power factor PF$ = \sigma S^2$ and/or to reduce the lattice thermal conductivity \cite{sales1996filled,fleurial1996highskutter,mandrus1997filledskutter,nolas2001recent,uher2001skutterudites,sales1999atomicClathrates,sales1998electroncrystal,wood1988materials,toberer2007local,chalfin2007cation,poudeu2006highthermoelectric,venkatasubramanian1997mocvd,venkatasubramanian2001thinfilm,beyer2002pbtesuper,lee1997thermalSiGe,Hsu2004cubicAg,caylor200enhancedPbTe5}. 
Increasing PF is challenging since $\sigma$ and $S$ have typically opposite dependence on the carrier concentration \cite{snyder2008complex}. It has been proposed that PF can be improved by several band engineering strategies. Among them is the concept of impurity resonant levels, close in energy to the Fermi level, that sharply increase the density of states near the resonant energy and are thought to enhance $S$ \cite{mahan1996thebest,he2017advances,Heremans2008enhancement,androulakis2010thermoelectric,heremans2012resonant}. Another is the idea of band convergence, where achieving high band degeneracy near conduction or valence band edges is discussed as 
beneficial for increasing $\sigma$ for a given $S$ due to the presence of multiple conducting
channels \cite{he2017advances, pei2011convergence, pei2011stabilizing, tang2015convergence, pei2012thermopower, liu2012convergence, lalonde2011leadtelluride, disalvo1999thermoelectric, mao2018advances, fu2014highBand,zhang2014highperformance,gibbs2017effective,xin2018valleytronics}. These concepts are actively pursued but are not fully investigated from the microscopic theoretical point of view.

The quantitative analysis of the band structure effects on the electronic transport properties has been enabled by first-principles calculations within the Boltzmann transport framework. In particular, the constant relaxation time approximation (CRT)\cite{madsen2006boltztrap}, where the electron scattering lifetime $\tau$ is taken to be a constant value, has been widely applied to estimate thermoelectric properties \cite{zhang2017discovery, xi2016bandengineering, chen2013importance, parker2015benefits, may2009influence,singh1997calculated}.
This approach is simple and often yields a reasonable approximation for $S$ (which is independent of $\tau$ if the latter is constant).
However, CRT approximation leaves the electron lifetime factor $\tau$ in the electrical conductivity $\sigma$ undetermined and requires an empirical fit. In recent years it has become clear that for predictive computational estimates of conductivity and PF it is necessary to take into account the electron lifetimes due to phonon and impurity scattering  \cite{samsonidze2018accelerated,Xi2018,witkoske2017thermoelectric,norouzzadeh2016classification,graziosi2020material}. First-principles calculations of electron-phonon coupling enabled the computation of intrinsic conductivity \cite{Ponce2020firstprinciples, giustino2007electronphonon, ponce2018towards, samsonidze2018accelerated, bang2018epamls, wee2019quntification, Deng2020epicstar} in good agreement with experiment and even led to the identification of new materials with high thermoelectric performance. Among them are half-Heusler NbFeSb and TaFeSb \cite{patent2013samsonidze, samsonidze2018accelerated} for which high $ZT$ values have been already confirmed experimentally  \cite{joshi2014nbfesb,Zhu2019discovery}.

While the necessity of computing $\tau(E)$ has been recognized for estimating the conductivity, much less has been discussed in the context of the Seebeck coefficient $S$. The classic approximate Mott expression for $S$ actually anticipates non-trivial contributions from the energy dependence of the carrier lifetime $\tau(E)$ in addition to the effect of the electronic density of states (DOS) $N(E)$ \cite{may2009influence}: 
\begin{equation}
\label{eq:Seebeck}
    \begin{aligned}
    S=-\frac{\pi^2k_BT}{3e}\left[\frac{\partial\ln(\sigma(E))}{\partial E}\right]_{E_f} \\
    = -\frac{\pi^2k_BT}{3e}\left[\frac{\partial\ln(N(E))}{\partial E}+
    \frac{\partial\ln(\tau(E))}{\partial E}\right]_{E_f} \;,
    \end{aligned}
\end{equation}
where $k_B$ is the Boltzmann constant, $e$ is the electron charge and $E_f$ is the Fermi level.
If the first term in Eq. \ref{eq:Seebeck} is dominant, the CRT approximation is sufficient, but when $\tau(E)$ has strong energy dependence, this approximation can fail qualitatively. 
So far significant contribution to $S$ in semiconductors from the second term of Eq. \ref{eq:Seebeck} has been discussed primarily in the context of scattering dominated at low temperatures by localized impurities. For instance, impurity scattering was hypothesized to be responsible for the measured change in the sign of $S$ in Ni-doped CoSb$_3$ below 20 K \cite{sun2015largeseebeck}. Energy-dependent scattering from resonant impurities was studied with an idealized Green's function model to suggest a possible enhancement and sign reversal of $S$ as well as a peak in $\sigma$ \cite{thebaud2019largeenhancement}. Abrupt changes of mobility with energy and temperature have been associated with observations of giant enhancements of $S$ values in materials near phase transitions, such as double perovskites\cite{Roy2018}, CuAlO$_2$ \cite{Mahmood2018} and  Cu$_2$Se\cite{Liu2013}, although it remains uncertain whether critical phenomena are responsible for these effects. An explanation of the large enhancement of $S$ in CuFeS$_2$ below 50 K was proposed by considering strong energy dependence of hopping transport \cite{Xie2020}.

In this work we introduce the possibility of engineering large PF enhancement and anomalous reversal of the Seebeck effect in semiconductors arising from intrinsic electronic and vibrational structure, in the absence of impurities and phase transitions. We consider the effects of the electron-phonon interaction on the electronic transport properties, since it is the dominant scattering mechanism in a material above its Debye temperature and, therefore, most relevant to consider in the context of potential applications for thermoelectric power generation. Using first-principles analysis of interband electron-phonon scattering, we find that the co-occurrence of two bands close in energy and with different effective masses leads to a strong dependence of $\tau(E)$ on the carrier energy $E$ near the Fermi level. This strong dependence drives the emergence of several anomalous trends in transport characteristics as functions of carrier concentration: non-monotonic dependence of conductivity $\sigma$ exhibiting a peak, reversal of the sign of the Seebeck coefficient $S$ relative to the majority carrier charge, and the appearance of a second peak in the power factor PF. These effects are found to be robust to temperature variations and high doping concentrations. We show that a two-band model with a simple estimate of $\tau(E)$ is sufficient to understand these effects, highlighting that interband scattering is in general responsible for the anomalous behavior, that emerges when a sharp change in the density of states occurs in the vicinity of the Fermi level. We propose a general microscopic understanding of this band engineering paradigm and systematically examine the electronic structure features that give rise to the transport anomalies. We illustrate these phenomena using as examples leading thermoelectric half-Heusler alloys TaFeSb and ZrNiSn and discuss the experimental considerations of designing materials in this regime. Finally, we re-examine key implications of the energy-dependent interband scattering for the band-convergence design strategy for thermoelectrics.

\section{Electronic transport calculations}
\label{sec:BTE}

In this work we model the electronic transport properties using the Boltzmann transport equation within the relaxation time approximation \cite{ziman1960EP}.
We employ first-principles calculations to obtain the energy-dependent electron lifetimes  $\tau(E)$ due to electron-phonon scattering.
These calculations use the electronic band structure approximated with density functional theory (DFT) \cite{hohenberg1964inhomogeneous,kohn1965selfconsistent} and vibrational spectra calculated using density functional perturbation theory (DFPT) \cite{baroni2001phonons}. A fully \textit{ab initio} description of electron-phonon scattering 
can be obtained starting with DFPT calculations of the electron-phonon coupling matrix elements. To compute $\tau(E)$, we use the electron-phonon averaged (EPA) approximation\cite{samsonidze2018accelerated, bang2018epamls,samsonidze} and validate it with the electron-phonon Wannier interpolation (EPW) method \cite{ giustino2007electronphonon, noffsinger2010epw, ponce2016epw,ponce2018towards, giustino2017electronphonon}. The key idea of the EPA method is to replace the expensive momentum-space integration of scattering contributions over the Brillouin zone in EPW by analytical integration over energies. The resulting EPA lifetime of a carrier with energy $E$, chemical potential (Fermi level) $\mu$ and at temperature $T$ is given by
\begin{equation}
\label{eq:EPAtau}
    \begin{aligned}
    \tau^{-1}(E,\mu,T)=\frac{2\pi\Omega}{g_s\hbar}\sum_{\nu}\bigl\{g_{\nu}^2(E,E+\bar{\omega}_{\nu})  [n(\bar{\omega}_\nu,T) \\ +f(E+\bar{\omega}_{\nu},\mu,T)]    \rho(E+\bar{\omega}_{\nu})  +g^2_{\nu}(E,E-\bar{\omega}_{\nu}) \\ \times \left[n(\bar{\omega}_{\nu},T)+1-f(E-\bar{\omega}_{\nu},\mu,T)\right]\rho(E-\bar{\omega}_{\nu})\bigr\} \;,
    \end{aligned}
\end{equation}
where $\rho(E)$ is the electronic DOS (number of electronic states per unit energy and unit volume), $f$ is the Fermi--Dirac distribution, $n$ is the Bose--Einstein distribution, $\bar{\omega}_{\nu}$ is the average energy of the phonon branch $\nu$, $\Omega$ is the unit cell volume, and $g_s=2$ is a spin factor in non-magnetic materials. Electronic transport coefficients can be defined in terms of the tensor $K^{(p)}$, which is defined as
\begin{equation}
    \label{eq:EPAtranspCoef}
    \begin{aligned}
    K^{(p)}_{\alpha\beta}(\mu,T)=\frac{g_se^{2-p}}{(2\pi)^3(k_BT)^{p+1}}\int dE v^2_{\alpha\beta}(E)\rho(E) \\ \times \tau(E,\mu,T)I^{(p)}(E,\mu,T) \;,
    \end{aligned}
\end{equation}
where $v^2_{\alpha\beta}(E)$ is the energy-projected squared velocity tensor,  $I^{(p)}(E,\mu,T) = (E-\mu)^p f(E,\mu,T) [1-f(E,\mu,T)]$, and the Greek letters $\alpha$ and $\beta $ denote Cartesian directions.
These coefficients $K_{\alpha\beta}^{(p)}(\mu,T)$ are directly related to electronic transport coefficients; namely, the electrical conductivity $\sigma$ is $\sigma_{\alpha\beta}(\mu,T)=K^{(0)}_{\alpha\beta}$, and the Seebeck coefficient $S$ is $S_{\alpha\beta}(\mu,T) = k_B \big[ \left(K^{(0)}\right)^{-1} \cdot K^{(1)} \big]_{\alpha\beta}$. In order to identify non-trivial effects of $\tau(E)$ on the behavior of the electronic transport properties as functions of carrier concentration and temperature, we compare the corresponding EPA and EPW predictions to those obtained using the CRT approximation. 
Further details on the computational methodology can be found in Sec. II of the Supplementary Material.

\begin{figure*}
    \centering
    \includegraphics[width=0.97\linewidth,trim=0.9cm 0cm 0cm 0cm]{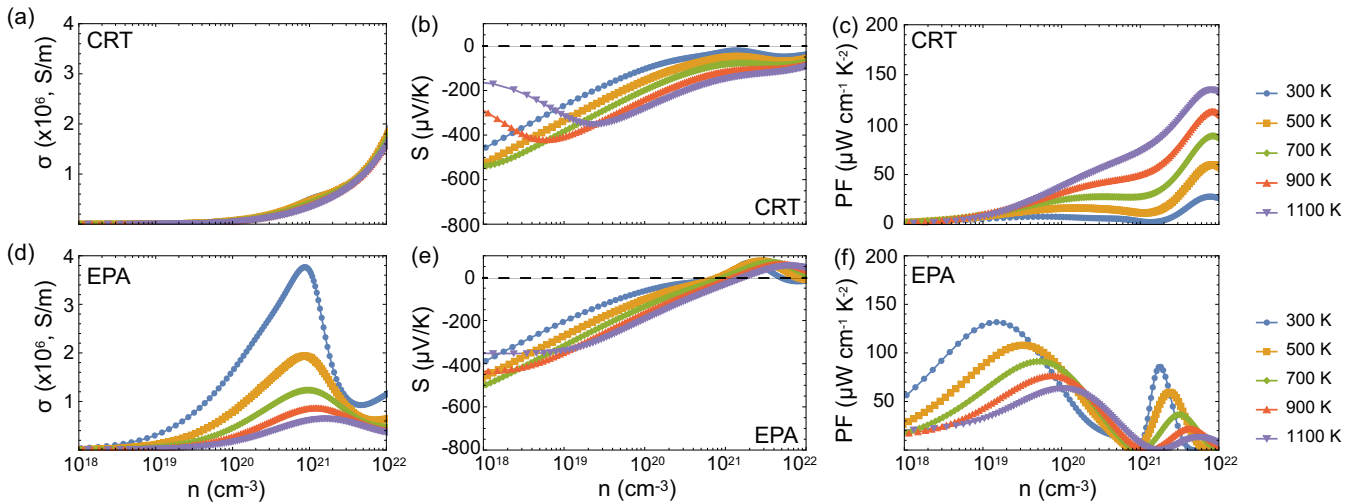}
    \caption{Electronic transport properties as functions of carrier concentration $n$ calculated for n-type TaFeSb using CRT (a-c) and EPA (d-f) approximations. 
    Panels (a) and (d) show the electrical conductivity $\sigma$, (b) and (e) - the Seebeck coefficient $S$, (c) and (f) - the power factor PF. 
    Each line color corresponds to a different temperature $T$ in the range 300 K to 1100 K.}
    \label{fig:crt_epa_tafesb}
\end{figure*}

\section{Transport properties of half-Heusler T\lowercase{a}F\lowercase{e}S\lowercase{b} and Z\lowercase{r}N\lowercase{i}S\lowercase{n}}
\label{sec:el_transport}
Half-Heusler (HH) alloys are a class of materials characterized by good thermal and mechanical stability, high power factors and, therefore, are attractive for applications for power generation at high temperatures \cite{he2017advances,  chen201recent3, zhu2018discoveryZrcobi, Fu2015realizing, zhu2015highefficiency,graf2011simpleRules}. 
Motivated by these considerations, computational screening of a wide range of compositions \cite{samsonidze2018accelerated, patent2013samsonidze} predicted p-type TaFeSb to be among the best HH thermoelectrics with $ZT\approx1$ at 673 K.
The p-type TaFeSb has been synthesized afterwards and its high $ZT$ has been experimentally confirmed \cite{Zhu2019discovery}. Interestingly, the same computational investigation predicted n-type TaFeSb to have an even higher $ZT$ upper limit value than p-type TaFeSb and higher than that of state-of-the-art n-type HH systems $M$NiSn ($M=$ Hf, Zr, Ti). This, however, has not been verified experimentally as of today to the best of our knowledge. A possible reason for this can be the narrow range of doping concentrations for which the optimal power factor was computationally obtained. In this section we analyze in detail the dependence of electronic transport properties of TaFeSb on carrier concentration and temperature and compare them with those of another state-of-the-art HH thermoelectric ZrNiSn. Our analysis below suggests that the mechanism responsible for the narrow carrier concentration range of high performance is related to anomalous features in the transport coefficients that emerge from inter-band electron-phonon scattering.

\begin{figure*}
    \centering
    \includegraphics[width=0.95\linewidth,trim=0.9cm 0cm 0cm 0cm]{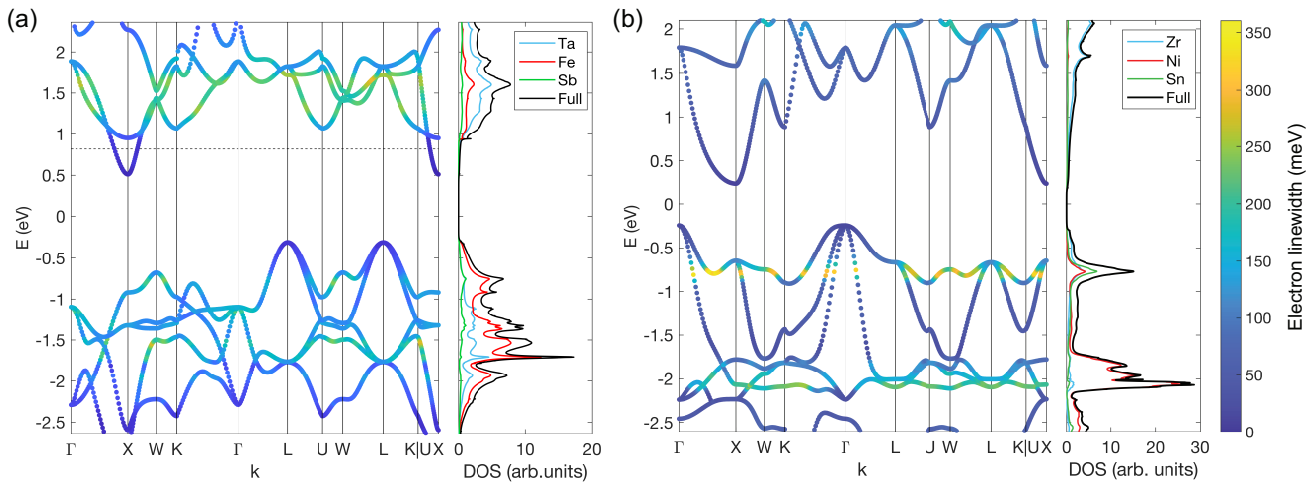}
    \caption{Electronic band structure and density of states of (a) TaFeSb and (b) ZrNiSn, referred to the Fermi level at 0 eV. 
    The color scale corresponds to the electron-phonon linewidth at 700 K, with the Fermi level (dashed line) of TaFeSb set in correspondence to the doping concentration at which a drop of the electrical conductivity occurs, i.e. $n\approx10^{21}$ cm$^{-3}$.}
    \label{fig:band_structure}
\end{figure*}

We start analyzing the electrical conductivity $\sigma$ of n-type TaFeSb. Fig. \ref{fig:crt_epa_tafesb} (a), shows $\sigma$  as predicted by the CRT approximation has a conventional trend of increasing with doping concentration and a weak decreasing trend with increasing temperature.
In contrast, $\sigma$ at the EPA level, shown in Fig. \ref{fig:crt_epa_tafesb} (d), exhibits a much stronger temperature dependence. Additionally, $\sigma$ has an optimal doping concentration of approximately $n=10^{21}$ cm$^{-3}$, largely independent of temperature, at which the electrical conductivity is maximized, a feature that is absent at the CRT level and is surprising in the conventional picture of doped semiconductors. As verification, Supplementary material (Fig. 2 (a)) shows that $\sigma$ obtained using the full momentum-space scattering integrals based on the EPW method has very similar behavior.

In conventional semiconductor physics, the sign of the Seebeck coefficient $S$ indicates the doping type of a semiconductor, i.e. $S<0$ for n-type doped semiconductors, and $S>0$ for p-type. This behavior is in accordance with the first DOS-dependent term of the Mott formula (Eq. \ref{eq:Seebeck}). We note that the situation in metals is less clear, where $S$ values are much smaller and the signs vary. It has been shown by first-principles calculations that phonon-limited electron lifetimes can explain the positive sign of $S$ in Li metal \cite{xu2014positiveSeebeck} and suggest a possible enhancement of the Seebeck coefficient due to localized $f$-electrons in YbAl$_3$ \cite{Liang2017}. The example of TaFeSb illustrates the general finding that intrinsic electron-phonon scattering can cause sufficiently strong energy variation of electron lifetimes to modify the conventional understanding of the Seebeck effect in semiconductors. As shown in Fig. \ref{fig:crt_epa_tafesb} (b), $S$ at the CRT level, considering the electronic DOS, is predicted to be always negative for n-type TaFeSb, as per conventional expectations. EPA results, presented in Fig. \ref{fig:crt_epa_tafesb} (e), show that $S$ changes sign, becoming positive at large-enough n-type doping concentrations. In Supplementary material (Fig. 2 (b)), the EPA results are validated with the EPW approach, showing qualitative and quantitative agreement.

The change in sign of $S$ deviates from conventional expectation, implying that the electronic response of n-type TaFeSb to an applied temperature gradient is typical of a p-type semiconductor, with holes as the majority carrier. To our knowledge, this is the first report of this anomaly in heavily doped semiconductors being caused by intrinsic electron-phonon scattering. Interestingly, $S$ changes sign at a similar doping concentration at which $\sigma$ begins decreasing, hinting at the connection between the phenomena. We also note that the Hall coefficient, computed at the EPA level of approximation, does not change sign (see Fig. 5 (c) of Supplementary Material) and therefore the magnetic response of the system remains that of an n-type conductor.


The power factor PF=$\sigma  S^2$ predictions using CRT and EPA approximations are shown in Fig. \ref{fig:crt_epa_tafesb} (c) and (f), respectively. CRT predicts a steady increase of the PF with doping concentration. In contrast, the EPA PF displays two peaks whose magnitude tends to decrease as the temperature is increased, while the presence of both peaks is robust to temperature changes. The emergence of two distinct peaks is a direct result of the anomalous properties of $\sigma$ and $S$ discussed above.
The change of the sign of $S$ implies that for some doping concentration, the PF drops to zero, before rising again. However, since the reversal in $S$ appears in near correspondence to the peak in $\sigma$, the PF quickly grows to large values at doping concentrations slightly larger than the PF minimum. As a result, the second sharp "anomalous" peak in PF appears. The combined effect is the shift the first peak to much low doping concentrations, while the second peak is prominent but narrow. This circumstance may explain the difficulty in experimental optimization of carrier concentration and performance of n-type TaFeSb.

\begin{figure*}
    \centering
    \includegraphics[width=0.99\linewidth,trim=0.9cm 0cm 0cm 0cm]{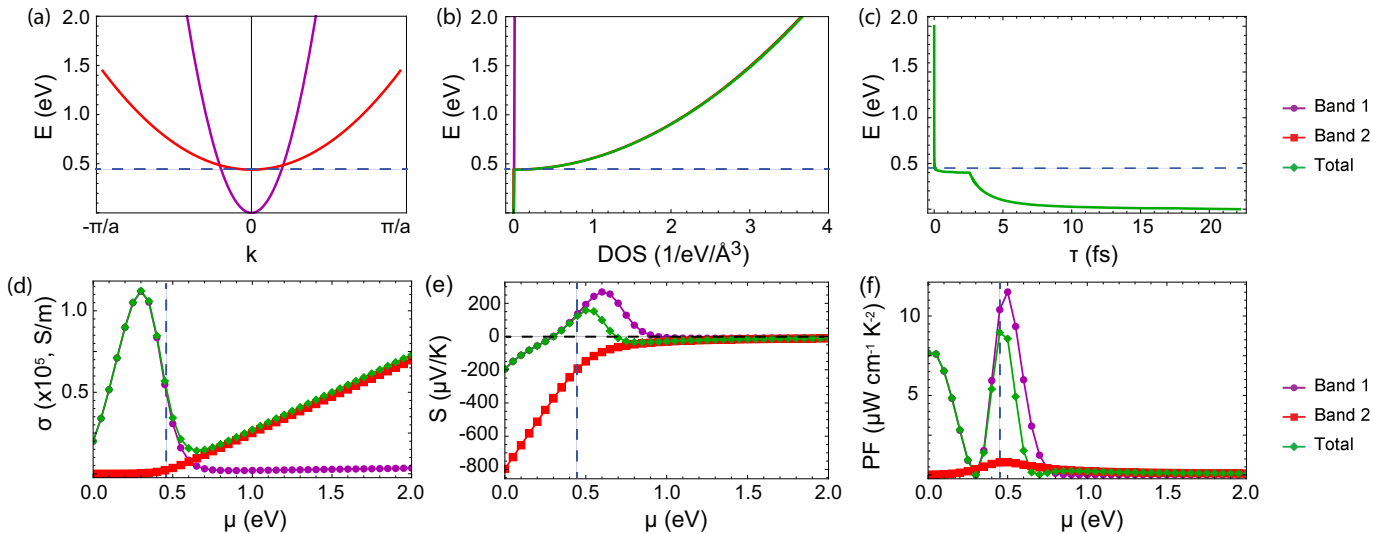}
    \caption{Results from a parabolic two-band model for n-type TaFeSb at fixed temperature of 700 K.
    Panel (a) shows the band structure of the model; (b) electronic density of states and its contributions from the two bands; note that the heavy band causes a step in DOS. 
    (c) energy dependence of the electron lifetime $\tau(E)$ for Fermi level placed exactly at the minimum of the higher energy band (both bands share the same values of $\tau$).
    Transport properties, i.e. electrical conductivity $\sigma$ (panel d), Seebeck coefficient $S$ (panel e) and power factor PF (panel f) as functions of Fermi level $\mu$.
    The model qualitatively reproduces the results observed in the full \textit{ab initio} calculations, namely a peak in $\sigma$ and a change in sign of $S$ at energies slightly below the second band minimum, and two peaks in PF, the second of which appearing at the energy for which $\mu$ matches the minimum of the higher-energy band, shown in the blue-dotted line.
    }
    \label{fig:tafesb_model}
\end{figure*}

For comparison, in the Supplementary Material (Fig. 3 (d-f)) we present the electronic transport coefficients obtained for p-type TaFeSb using the EPA approximation. We do not observe any anomalous characteristics in the transport properties in the wide range of doping concentrations. We note that CRT still displays notable differences compared to the EPA, predicting a weak dependence of $\sigma$ on temperature and a different  doping concentration for optimal PF. While p-type TaFeSb does not present anomalous behaviors of transport coefficients, its study  allows us to validate computational predictions against experimental measurements of Ref. \onlinecite{Zhu2019discovery}, (see Fig. 4 of Supplementary Material). Our predictions for $S$ are in quantitatively good agreement with experimental measurements. We see that $\sigma$ is qualitatively well predicted at the EPA or EPW level; however, both approaches overestimate $\sigma$, especially at low $T$. This is not surprising, especially below the Debye temperature, since the electron scattering due to impurities, grain boundaries, and alloy disorder are present in the samples but neglected in our simulations. Additional sources of disagreement may include uncertainties in the experimentally determined carrier concentrations, composition differences from pure TaFeSb, and the inaccuracy of DFT band structures.

As another example, we also analyze the electronic transport properties of n-type and p-type ZrNiSn. 
Here we only summarize the two most important results, while details are found in Sec. III B of the Supplementary Material. First, n-type ZrNiSn qualitatively behaves similarly to p-type TaFeSb, with transport coefficients following trends expected in conventional semiconductors: $\sigma$ decreases with temperature and increases with doping concentration, $S$ is negative and PF has a single peak.
Second, p-type ZrNiSn shares the same anomalous transport characteristics of n-type TaFeSb, with a maximum of $\sigma$ at carrier concentration close to where $S$ reverses its sign, and the presence of two peaks in the PF. As discussed previously, only EPA and EPW methods produce this behavior while the CRT approximation does not, confirming the role of electron-phonon scattering, rather than band structure alone, in causing these transport phenomena.

\section{Physical origins of the transport anomalies}
\label{subsec:origin_anomalous}

In this section, we explain the mechanism at the origin of the anomalous transport trends observed above, and establish the necessary physical features for their appearance. 
We start by analyzing the bottom of the conduction bands of TaFeSb, shown in Fig. \ref{fig:band_structure}(a), which we plot together with the electron linewidth, defined as \cite{ponce2018towards,giustino2017electronphonon} $\Gamma_{n\boldsymbol{k}} = 1/\tau_{n\boldsymbol{k}}$, obtained using electron-phonon contributions.
The conduction band edge of TaFeSb consists of a single non-degenerate parabolic band at the $X$ point, where linewidths take the smallest values.
There are additional subvalleys at an energy of approximately 0.5 eV above the conduction band minimum: a three-fold degenerate heavy-band subvalley at the $X$ point, and 12-fold degenerate light-band subvalleys at the $K$ and $U$ points.
We also note that linewidths at $U$ and $K$ subvalleys are much larger than at the $X$ point.
The heavy-band subvalleys produce a sharp increase in the electron DOS, also shown in Fig. \ref{fig:band_structure}(a).
The $X$, $K$, and $U$ subvalleys appear relevant, since their energy levels correspond to the doping concentration at which the anomalies take place, such as the sign reversal of $S$.

To analyze the physical mechanisms, we propose to model the band structure of TaFeSb with a two-parabolic-bands model, consisting of a light and a heavy band crossing each other.
The two parabolic bands are offset vertically, as shown in Fig. \ref{fig:tafesb_model} (a).
In this model, the band dispersion relation is $E_{i}(\boldsymbol{k}) = \frac{\hbar^2 k^2}{2 m_{i}} + \Delta E_{i}$, where $m_{i}$ is the effective mass of the light ($i=1$) or heavy ($i=2$) band, and $E_{0,i}$ is the band offset.
The parameters are fitted to the conduction band minimum and subvalleys of DFT band structure of TaFeSb, specifically $m_1=0.59 m_{\bar{e}}$, $m_2=6.32m_{\bar{e}}$, $\Delta E_{1}=0$ eV and $\Delta E_{2}=0.44$ eV.
In this simplified model, the DOS  is given by $ \rho_{i}(E) = \sqrt{2}g_s (2\pi^2\hbar^3)^{-1} N_{tot,i}   m_{i}^{3/2} \sqrt{E-\Delta E_{i}}  $, where $g_s$ is the spin degeneracy ($g_s=2$ for spin-unpolarized calculations as considered here), and the total band degeneracy $N_{tot,i}$ consists of orbital degeneracy $N_{o,i}$ and valley degeneracy  $N_{v,i}$, so that $N_{tot,i}=N_{o,i} \times N_{v,i}$. 
The total DOS, presented in Fig. \ref{fig:tafesb_model}(b), is small at low energies, where only the light band contributes. At energy higher than $\Delta E_{2}$ the contribution from the heavy band dominates the total DOS, also due to its large degeneracy. 

\begin{figure*}
    \centering
    \includegraphics[width=1\linewidth,trim=0cm 0cm 0cm 0cm]{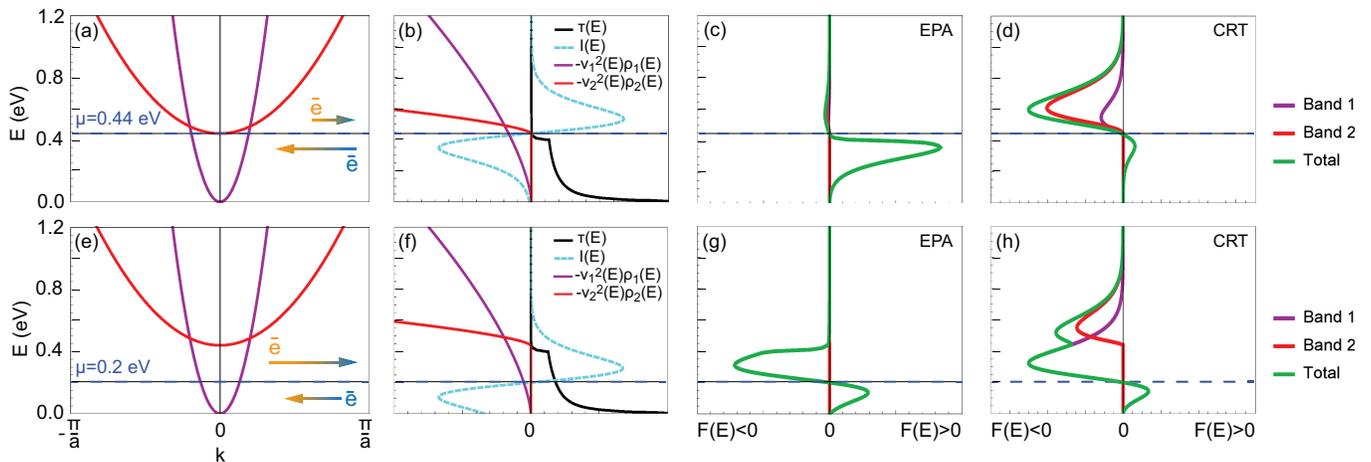}
    \caption{
    Analysis of the change of sign in Seebeck coefficient $S$ using the two-parabolic-bands model of n-TaFeSb at 700K.
    Top row: analysis of $S$ when the Fermi level $\mu$ is at the minimum of the higher-energy band ($S>0$). 
    Bottom row instead analyzes the case where the Fermi level lies below the second band minimum ($S<0$).
    Panels (a) and (e) show the model band structures, with $\mu$ shown by the dashed lines.
    Arrows demonstrate the direction of the electron diffusion: for $E>\mu$ - from hot to cold end of device, for $E<\mu$ - from cold to hot end. The arrow size indicates which electron flux dominates. 
    Panels (b) and (f) show the different contributions to $S$, as defined in the main text. 
    Note in particular that the sharp increase in the density-of-state projected velocity $v^2 \rho$ at $E\approx0.5$ eV is counterbalanced by a sharper decrease in the lifetime $\tau$.
Energy and band resolved contributions to the Seebeck coefficients $F_1(E)$, $F_2(E)$ and $F(E)$ calculated using the EPA approximation are shown in panels (c) and (g), and using the CRT approximation in panels (d) and (h).}
    \label{fig:seebeck_expl}
\end{figure*}

To include the energy dependence of lifetimes $\tau(E)$ in our model, we start from the EPA estimate (Eq. \ref{eq:EPAtau}), and further simplify it by setting the electron-phonon coupling to a constant ($g^2=0.004$ eV$^2$), motivated by the observation that it does not vary significantly with energy \cite{samsonidze2018accelerated}.
The average phonon frequencies are chosen using the \textit{ab initio} calculations (see Sec. IVA and Fig. 9(a) of the Supplementary Material).
Moreover, we neglect the impact of selection rules and assume that all scattering processes are allowed, provided energy is conserved.
By doing this, we implicitly allow for the presence of both intraband and interband scattering. The code with model implementation is provided in the Supplementary Material.
In Fig. \ref{fig:tafesb_model}(c), we show the estimated energy-dependent electron lifetime.
At low energies ($E<\Delta E_{2}=0.44$ eV), only electrons in the light band are in play, and the lifetime increases rapidly as it reaches the minimum of the conduction band.
At high energy, we observe a sharp drop of lifetimes, following the expected inverse relationship between $\tau(E)$ and the DOS associated with the phase-space available for electron-phonon scattering.

We now show that this simple model is sufficient to capture the anomalous features of the electronic transport coefficients in n-type TaFeSb. The electrical conductivity $\sigma$, shown in Fig. \ref{fig:tafesb_model}(d) as a function of the Fermi level $\mu$, exhibits a non-monotonic trend, as in the full \textit{ab initio} simulations. The decrease of $\sigma$ from its peak value starts slightly below the point where $\mu$ reaches the heavier band, and there is an inflection point in correspondence of the minimum of the heavy band.
Decomposing the conductivity into contributions from each band, we see that the electrical conductivity at lower energies is fully determined by the light band 1, with contributions from the heavier band dominating at energies above the heavy band 2 minimum $\Delta E_{2}$.
This allows us to understand the role of interband electron-phonon scattering: in a single-parabolic-band model, conductivity would keep increasing as the Fermi level shifts deeper into the conduction band.
However, this is not the case here: as the Fermi level shifts to include carriers in the heavier band, electrons from the light band can scatter against heavy electrons, suppressing their lifetimes and thus $\sigma$ (see also Fig. 12 of the Supplementary Material for more details).
At higher energies, $\sigma$ is largely determined by the heavy band but, since electrons are slower, $\sigma$ results smaller than for lower doping concentrations.
We stress that this result cannot be reproduced with a CRT approach: when $\tau$ is constant, the sharp decrease of $\tau(E)$ shown in Fig. \ref{fig:tafesb_model}(c) is not captured and, as a result, the drop of $\sigma$ at 0.44 eV does not occur (see Fig. 10(a) and 12 of The Supplementary Material). Clearly, this behavior also does not occur, if only intraband scattering is taken into account (see Fig. 10(d) of the Supplementary Material).
We note in passing that Eq. \ref{eq:EPAtranspCoef} for $p=0$ explains the linear dependence of $\sigma$ observed at low $\mu$ or for $\mu\gg0.5$ eV. 
In fact, for a single parabolic band model, the DOS scales as $\sqrt{E}$, but this dependence is canceled by the $1/\sqrt{E}$ dependence of the lifetimes; the diagonal components of the electron velocity scale as $v^2(E)=(\frac{\partial E}{\partial k})^2 \propto E$, and the term $I(E)$ merely selects a value of energy close to $\mu$.
As a result, $\sigma$ is expected to grow linearly with $\mu$ for values of energy far from the intersection of the two bands.

\begin{figure*}
    \centering
    \includegraphics[width=0.99\linewidth,trim=0.9cm 0cm 0cm 0cm]{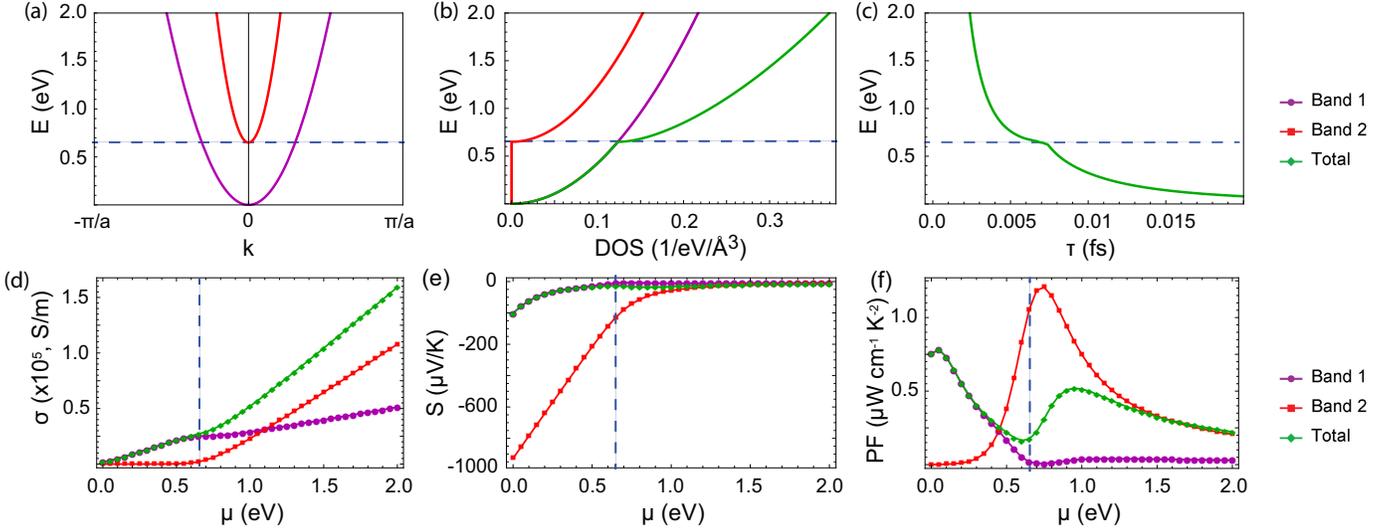}
    \caption{Two-parabolic-bands model of transport for n-type ZrNiSn at temperature of 700 K. 
    Panel (a) shows the electronic band structure, where the higher-energy band is now lighter. 
    The Fermi level (blue dashed line) is set at the heavy second band minimum.
    (b) electronic density of states, where we note that the DOS step is less pronounced than for n-type TaFeSb.
    (c) energy dependence of the electron lifetime $\tau$, that, like the DOS, has a smoother behavior than for n-type TaFeSb.
    Finally, the electrical conductivity $\sigma$, the Seebeck coefficient $S$ and the power factor PF are shown as functions of Fermi level in panels (d), (e) and (f), respectively.
    At variance with TaFeSb, where the second band is heavier than the first, the peak in $\sigma$ is suppressed, $S$ doesn't reverse sign, and the second peak in PF appears washed out.
    }
    \label{fig:zrnisn_model}
\end{figure*}

The two-band model also successfully captures the sign reversal of the Seebeck coefficient $S$, which happens slightly below the bottom of the heavier band at $\mu\approx0.4$ eV, as shown in Fig. \ref{fig:tafesb_model}(e). The sign change originates from a competition between the two bands, which can be better understood with the aid of Fig. \ref{fig:seebeck_expl}. 
In detail, we study the energy dependence of the factors entering the integrand $F(E)$ of the transport coefficient $K^{(1)}$ (see Eq. \ref{eq:EPAtranspCoef}) which determines $S$. 
The top row in Fig. \ref{fig:seebeck_expl} panels (a-d) display the situation when the Fermi level is at the minimum of the heavy band ($\mu=\Delta E_{2}=0.44$ eV) at temperature T=700 K, which results in a anomalous positive Seebeck coefficient.
Fig. \ref{fig:seebeck_expl}(b) examines how the various components contribute to the coefficient $K^{(1)}$ as a function of energy.
The function $I(E)$ acts as a filter that activates only electrons within a few $k_BT$ of the Fermi level.
$I(E)$ has opposite signs above and below $\mu$, indicating that the higher energy electrons and lower energy electrons respond differently to applied temperature gradient: high energy electrons diffuse from hot to cold side of the thermoelectric sample and low energy electrons diffuse from cold to hot side.
Electron lifetimes, which are roughly proportional to the inverse of DOS, are large for $E<\mu$ and negligibly small for $E>\mu$, where they are suppressed by interband scattering.
The factor $-v^2(E)\rho(E)$ is shown in Fig. \ref{fig:seebeck_expl}(b) in terms of the contributions from each band.
The heavy band $-v_2^2(E)\rho_2(E)$ gives a large contribution, but only at energies $E>\mu$, so that at small energies only the light band contributes.
Combining these quantities together, the sign of $S$ results from a competition between electrons above and below $\mu$, due to the anti-symmetry of $I(E)$. 

To see which term dominates, in Fig. \ref{fig:seebeck_expl}(c) we plot $F(E)$, the integrand resulting from multiplying the factors discussed above, as well as the individual band contributions $F_1(E)$ and $F_2(E)$.
We see that the contributions from high-energy electrons $E>\mu=\Delta E_{2}$ are suppressed by the lifetime reduction due to interband electron-phonon scattering. As a result, the system behaves as a "p-type" thermoelectric ($S>0$), despite being strongly n-doped. This follows from the fact that lifetimes of both bands are inversely proportional to the total DOS $\rho(E)$.
Fig. \ref{fig:seebeck_expl}(d) illustrates why the CRT approximation does not capture this behavior: with a constant $\tau$, no suppression of transport contributions from high energy electrons takes place and, given the large values of $-v^2(E)\rho(E)$ at higher energies, the Seebeck coefficient is predicted to be negative.

In the bottom row of Fig. \ref{fig:seebeck_expl}, panels (d-f) we repeat the analysis to understand the negative values of $S$ for values of $\mu$ well below the heavy band (Fig. \ref{fig:seebeck_expl}(e)).
The only difference to the previous case is that the anti-symmetric factor $I(E)$ is now centered around the lower value of $\mu$, where it only allows contributions from the light band, with the heavy band not thermally occupied (see Fig. \ref{fig:seebeck_expl}(f)).
Since lifetimes at these low energies are not suppressed by interband scattering, both electrons above and below $\mu$ effectively contribute to $S$.
The overall integrand (Fig. \ref{fig:seebeck_expl}(g)) is now dominated by electrons with energy $E>\mu$ due to their higher DOS, resulting in an overall negative $S$.
We note that the CRT model captures the correct sign in this scenario (Fig. \ref{fig:seebeck_expl}(h)): approximating $\tau$ as a constant is a better approximation here, since the factor $I(E)$ filters out contributions from energies at which $\tau(E)$ has the largest variation.

\begin{figure*}
    \centering
    \includegraphics[width=0.8\linewidth,trim=0cm 0cm 0cm 0cm]{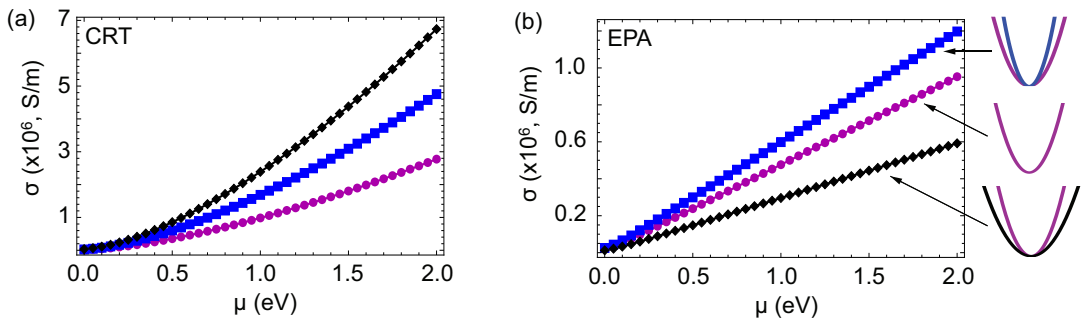}
    \caption{Change to electrical conductivity $\sigma$ upon adding a light or heavy band on top of an existing band with the same energy minimum, estimated at a temperature of 700 K and plotted as a function of the Fermi level $\mu$.
    In purple color, $\sigma$ for a system with a single band.
    In black and blue, the resulting $\sigma$ after adding a heavier or lighter band, respectively.
    Left panel shows the predictions from the CRT approximation. Right panel shows the estimates of EPA approximation using an energy-dependent lifetime.}
    \label{fig:convergence}
\end{figure*}

The two-band model also reveals why n-type ZrNiSn, for example, does not exhibit the anomalous features discussed for n-type TaFeSb.
In this case, we model the full band structure with two bands with effective masses $m_1=3.84m_e$ and $m_2=0.87m_e$, and offset of $\Delta E_{2}=0.65$ eV (additional parameters are found in Sec. IV B of the Supplementary Material).
This time, since the difference in band effective masses is smaller, the total DOS (Fig. \ref{fig:zrnisn_model}(b)) does not show an increase as steep as in the TaFeSb case.
As a consequence, high-energy electron lifetimes (Fig. \ref{fig:zrnisn_model}(c)) are not dramatically suppressed by inter-band scattering and no anomalous effects takes place. As the doping level $\mu$ is varied, the conductivity $\sigma$ (Fig. \ref{fig:zrnisn_model}(d)) does not show a peak, and $S$ does not change sign (Fig. \ref{fig:zrnisn_model}(e)).

We note that the above analysis does not qualitatively depend on the location of the two band minima in the Brillouin zone, we chose them to be at the same k-point for simplicity. This is because only energy-dependent quantities are included in the EPA transport formalism.

\section{Band convergence}
One of the strategies of using band engineering to design high-performance thermoelectric materials is based on the concept of band convergence, where it is proposed that high degeneracy near conduction or valence band edges is beneficial for increasing $\sigma$ for a given $S$ due to the presence of multiple conducting channels \cite{he2017advances, pei2011convergence, pei2011stabilizing, tang2015convergence, pei2012thermopower, liu2012convergence, lalonde2011leadtelluride, disalvo1999thermoelectric, mao2018advances, fu2014highBand,zhang2014highperformance,gibbs2017effective,xin2018valleytronics}. However, several recent studies predicted that the interband scattering can complicate the conventional understanding of the effects of band convergence on the electronic transport properties and that it is not always beneficial for the PF optimization  \cite{park2020high, kumarasinghe2019bandAlignment,park2020whenband}.  Here we revisit the concept of band convergence using the parabolic two-band model within the EPA approximation. We study a limiting scenario where two bands of different mass are at the same offset. This, for example, describes the top of the valence band of TaFeSb (at the $L$-point)  or ZrNiSn (at the $\Gamma$-point).
To understand the impact of interband scattering, we first consider a single parabolic conduction band \#1 of mass $m_1 = 0.59 m_e$ with no degeneracy ($N_1=1$).
Both the CRT and the EPA predict qualitatively similar trends, namely $\sigma$ increasing monotonically as the Fermi level shifts into the conduction band, as shown in purple color in Fig. \ref{fig:convergence}, for $T=700$ K.
As explained before, EPA shows a linear increase of $\sigma$ with $\mu$, resulting from the energy-dependence of the lifetimes on the DOS inverse ($\tau \sim 1/\sqrt{E}$).
The CRT approximation instead predicts a faster-than-linear increase of $\sigma$ with $\mu$, due to the missing energy-dependence of lifetimes.
Next, we add another single-degenerate ($N_2=1$) band \#2 with the same minimum (offset) as band \#1. 
We consider two scenarios, (i) when the added band \#2 is heavier than \#1, $m_2>m_1$ and (ii) the opposite case when $m_2<m_1$ (in detail, we set $m_{2,h}=1.19m_e$ for (i) and  $m_{2,l}=0.29m_e$ for (ii)). 
Intuitively, one would expect that adding a lighter band would increase conductivity, since a lighter carrier is expected to have a higher mobility.
This is indeed predicted at the EPA level: in the right panel of Fig. \ref{fig:convergence} we show that case (i) results in a lower $\sigma$  (shown as a black line) than for a single band \#1, as a result of additional interband scattering. In case (ii) the addition of the lighter band increases $\sigma$ (shown with the blue line).
However, the CRT approximation predicts opposite behavior of $\sigma$, with the addition of a heavier band resulting in larger conductivity as shown in the left panel of Fig. \ref{fig:convergence}.
This behavior is due to the fact that the CRT approximation overestimates the impact of the DOS on the conductivity, neglecting the decrease of the lifetime due to the interband scattering.
Therefore, it is crucial to consider the energy dependence of lifetimes. This effect is simplest to illustrate by varying the degeneracy of a single band. For example, consider a doubly degenerate band ($N=2$) which is identical to having two singly-degenerate bands with the same effective masses perfectly converged. The CRT approximation predicts doubling of the electrical conductivity and PF compared to the case of a single band ($N=1$) due to the increase in the DOS. If energy dependence of lifetime $\tau(E)$ is taken into account, e.g. with the EPA approximation, $\sigma$ and PF are independent of the degeneracy, since the enhancement of DOS $\rho(E)$ is compensated by the reduction of the electron lifetimes $\tau(E)\sim1/\rho(E)$. The Seebeck coefficient in both CRT and EPA models is independent of the degeneracy $N$ value.

In Figs. 13 and 14 of Supplementary Material we discuss other details of this case, such as estimates of $S$ and the impact of band degeneracy. 
We note that earlier studies have observed that adding a heavier band is detrimental to electronic conductivity \cite{kumarasinghe2019bandAlignment,park2020whenband}, in agreement with our conclusions.


\section{Materials design using interband scattering}
Using the simple parabolic two-band EPA model that successfully describes the transport characteristics, we now examine the possibility of using the interband scattering to design materials with high PF, i.e. what is the band structure that maximizes the anomalous PF peak at high carrier concentrations.

\begin{figure}
    \centering
    \includegraphics[width=1\linewidth,trim=0cm 0cm 0cm 0cm]{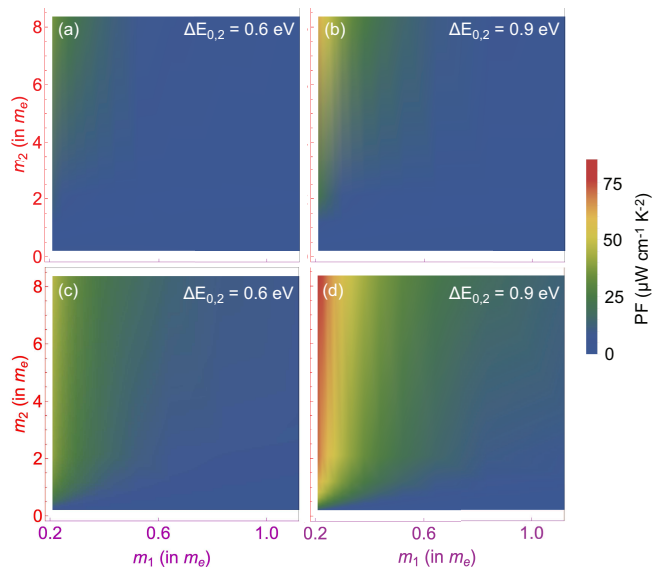}
    \caption{Magnitude of the anomalous second peak in the power factor as a function of band effective masses $m_1$ and $m_2$ in the parabolic two-band model. PF values correspond to the Fermi level placed exactly at the extremum of the band 2 ($\mu=\Delta E_{2}$). Panels (a) and (b) correspond to band degeneracies of $N_{tot,1}=3$ and $N_{tot,2}=1$; panels (c) and (d) correspond to band degeneracies of $N_{tot,1}=3$ and $N_{tot,2}=27$.}
    \label{fig:heatmaps}
\end{figure}

In Fig. \ref{fig:heatmaps}, we plot the magnitude of the anomalous second PF peak as a function of band effective masses $m_1$ and $m_2$ varied in the range from $0.21m_e$ to $8.36m_e$ ($m_1$ is only shown up to $1.1m_e$ since there is no effect at higher values). In panels (a) and (b) band degeneracies are $N_{tot,1}=3$ and $N_{tot,2}=1$, while we choose $N_{tot,1}=3$ and $N_{tot,2}=27$ for (c) and (d). The band offset in plots (a) and (c) is $\Delta E_{2}$=0.6 eV and 0.9 eV in (b) and (d). All other model parameters used in these calculations are fixed to those described in Sec. \ref{subsec:origin_anomalous} for n-type TaFeSb. In all these cases, the Fermi level is placed at the minimum of the second heavier band $\mu=\Delta E_{2}$.
More estimates of PF for other parameter values are shown in Fig. 15 of the Supplementary Material. 
From this Figure we can infer the conditions for which the anomalous peak in the PF is maximized: (i) the second higher-energy band is significantly heavier than the lower-energy light band $m_2\gg m_1$, (ii) the band  offset $\Delta E_{2}$ is large (thus requiring higher doping concentrations), and (iii) the second heavy band has large degeneracy $N_{tot,2}\gg N_{tot,1}$. 
These conditions can be rationalized in terms of the observations made in the Sec. \ref{subsec:origin_anomalous}.
First, small values of $m_1$ allow for large values of $\sigma$ at low doping concentrations when only the low-energy band contributes to transport.
Second, a large band offset $\Delta E_{2}$ allows larger values of the peak in  $\sigma$ as a function of $\mu$.
Third, a larger degeneracy of the heavy band allows for a stronger energy dependence (suppression) of lifetimes for $\mu > \Delta E_{2}$.
We therefore anticipate that it is possible to explore a wide space of materials to optimize thermoelectric performance in presence of the anomalous transport characteristics studied here.

\section{Summary and conclusions}

We showed that intrinsic interband scattering mediated by electron-phonon interaction in semiconductors can give rise to anomalous features of thermoelectric transport coefficients, at practically relevant doping levels and elevated temperature. Specifically, above a certain doping concentration, the electrical conductivity begins to decrease, the Seebeck coefficient changes sign relative to the Hall coefficient, and the power factor develops an additional peak. The root cause of these effects is the rapid variation of the electron lifetime as a function of energy, which occurs when bands of different curvature and multiplicity coexist at the same energy, allowing for electrons in the light band to be scattered to the heavy band. We used \textit{ab initio} calculations of electron lifetimes and the Boltzmann transport formalism to examine transport properties of high-performance half-Heusler compounds, demonstrating that these anomalies are captured only if the energy dependence of electron lifetimes is taken into account, and are absent in the commonly used constant relaxation time approximation. We also constructed a simple parabolic two-band model, based on the electron-phonon averaged approximation, that faithfully reproduces the main features of the anomalous transport behavior. 
We showed that when the two bands are close in energy, the rapid variation of electron lifetime with energy translates to the detrimental effect of heavy bands on transport characteristics. This observation revises the commonly discussed band-convergence strategy for designing high-performance thermoelectric materials.
Importantly, using the the two-band electron-phonon model, we identified specific band engineering design rules for maximizing the anomalous power factor. It was found that in the regime of sign reversal of the Seebeck effect, the power factor peak can be narrow, so that slight variations in the doping concentration may result in large variations of thermoelectric performance. The general understanding developed using the outlined modeling approach provides a simple strategy for finding new classes of unconventional semiconductor materials with high thermoelectric performance, using only the knowledge of the electronic band structure.


\section{Acknowledgments}
We would like to thank Georgy Samsonidze, Jennifer Coulter, Nakib Protik, Daehyun Wee, Semi Bang, and Jorge \'{I}\~{n}\'iguez for fruitful discussions.
N.S.F. thanks the Swiss National Science Foundation (project number P2EZP2\_178532) for financial support. B.K acknowledges support from the Harvard University Climate Change Solutions Fund.
Computational resources were provided by XSEDE (projects DMR190062 and DMR190077) and Harvard University Research Computing.
\bibliographystyle{apsrev}
\bibliography{main.bib}

\end{document}